\documentclass[twocolumn,preprintnumbers,amsmath,amssymb]{revtex4}

\usepackage{graphicx}% Include figure files
\usepackage{dcolumn}% Align table columns on decimal point
\usepackage{bm}% bold math
\newcommand{\mog}{MoS$_2$/G}
\newcommand{\mo}{MoS$_2$}
\newcommand{\lit}{Li$_{\rm T}$}
\newcommand{\lih}{Li$_{\rm H}$}
\newcommand{\g}{graphene}
\begin{document}
%\preprint{APS/123-QED}

\title{Lithium incorporation at the \mo/graphene interface: an {\it
ab initio} investigation}

\author{R. H. Miwa$^1$,  and W. L. Scopel$^2$}

\affiliation{$^1$Instituto de F\'{\i}sica, Universidade Federal de
  Uberl\^andia, C.P. 593,   38400-902, Uberl\^andia, MG, Brazil.\\ }

\affiliation{$^2$Departamento de F\'{\i}sica, Universidade Federal do Esp\'irito
Santo, 299075-910,  Vit\'oria, ES, Brazil \\ and Universidade Federal
Fluminense, 277255-250, Volta Redonda, RJ, Brazil, }

\date\today

\begin{abstract}

  Based on {\it  ab initio} calculations,  we examine the incorporation of Li
atoms in the \mo/graphene interface. We find that  the intercalated Li atoms are
energetically more stable than Li atoms adsorbed on the \mo\ surface. The
intercalated atoms interact with both graphene sheet and \mo\ layer,
increasing the Li binding energies. However, the equilibrium geometries are
ruled by the \mo\ layer, where the intercalated Li atoms lie on the  top (\lit)
and hollow (\lih) sites of the \mo\ layer.
% 
% Where the former configuration is energetically more
% stable by 0.1~eV compared with \lih. The same total energy difference was
% verified for Li adatom on the \mo\ surface, suggesting the same \lit
% $\rightarrow$ \lih\ jumping rate in the \mo/G interface and on the \mo\ surface.
We calculate the Li diffusion barriers, along the \lit $\rightarrow$ \lih\
diffusion path, where we find similar energy barriers compared
with that obtained for Li adatoms on the \mo\ surface.
Our results allow us to infer that the Li storage capacity increases
at \mo/G interfaces, in comparison with Li adatoms on the \mo\ surface, however,
with no reduction on the mobility of the intercalated Li atoms. Those
properties are interesting/useful to the development of Li batteries based on
\mo.

\end{abstract}

%\pacs{ ... }

\maketitle

%\newpage

%\begin{multicols}{2}

%\section{Introduction}

Two dimensional (2D) materials have attracted numerous studies addressing
technological applications as well as fundamental science. In the past few
years,  graphene has been the subject of the most of studies in those 2D
system~\cite{novoselovScience2004,novoselovNat2005}. Indeed, graphene exhibits a
strong technological appeal mainly to the development of new electronic
nano-devices~\cite{wangPRL2008}. On the other hand, (recently) other 2D layered
structures, composed by few atoms, have been considered as interesting/promising
materials for several applications.  In particular the molybdenum di-sulfide
(\mo) to the development of electronic
devices~\cite{radisavljevicNatNanotech2011,baoAPL2013,dasNanoLett2013}, and Li
batteries (LIBs)~\cite{duChemComm2010}. \mo\ is a hexagonal layered structure,
where each layer is composed by S-Mo-S atoms covalently bonded, stacked along
the [0001] direction, while the layer--layer interaction is weak, mediated by
van der Waals (vdW) forces. Such layered structure promotes the intercalation of
Li ions.

Very recently, epitaxial layers of \mo\ have been successfully synthesized on
the graphene sheet~\cite{shiNanoLett2012,leeNanoLett2013}, \mog, although the
large lattice mismatch between \mo\ and graphene. In this case, there are no
chemical bonds at the \mo/G interface, and the \mo\ layers are attached to the
graphene substrate through vdW interactions.  On the other hand, recent
experimental studies indicate that \mo/G composites present a set of quite
interesting/suitable electrochemical properties to the development of high
performace LIBs. The authors suggest that such high performance of LIBs, based
on \mog\ systems, is due to ``synergistic effects between layered MoS2 and
graphene''~\cite{changACSNano2011,changJMatChem2011,changChemComm2011}. In this
work we try to clarify the rule played by those ``synergistic effects`` to the
Li incorporation in \mog.

\begin{figure}[h]
%%%
% Fig 1
%%%
\includegraphics[width= 8.5cm]{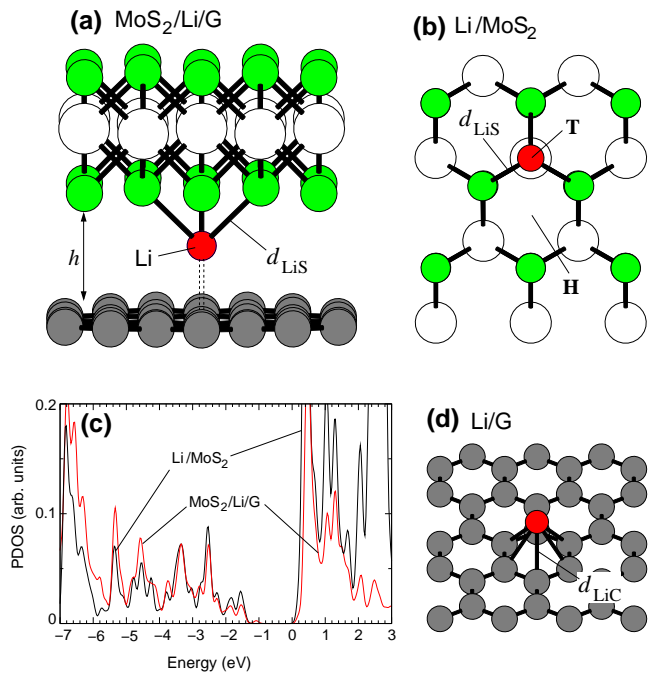}
\caption{Structural models of the Li intercalated \mo/Li/G (a), and 
the Li adsorbed Li/\mog\ (b) systems. (c) Projected density of states of \lit\
atoms for the Li/\mo\ (black-lines) and \mo/Li/G (red-lines) systems.
(d) Structural model of the Li/G system, for a Li adatom on the graphene H
site.}
\label{models}
\end{figure}

We have performed a theoretical {\it ab initio} study of Li atoms in the \mo/G
interface (\mo/Li/G). Initially we verify the energetic stability of the \mo/G
interface, and then the energetic properties of Li atoms adsorbed on the \mo\
surface (Li/\mo), and on the graphene sheet (Li/G). For the \mo/Li/G system, we
have considered a set configurations for the Li atoms in the \mo/G interface
region. Based on the calculation of the Li binding energies, we find an
energetic preference for the \mo/Li/G system, in comparison with the
Li/\mo\ and Li/G configurations. The energetically stable geometries of Li atoms
at the interface region are the same as those obtained for Li adatoms on the
\mo\ surface, namely,  Li atoms lying on the top site (\lit) aligned with the Mo
atom, and on the hexagonal hole site (\lih) of \mo. We find that the \lit\
configuration is energetically more stable than \lih\ by 0.1~eV for both
systems, \mo/Li/G and Li/\mo, thus suggesting  the same \lit\ $\rightarrow$
\lih\ jumping rate for the intercalated (\mo/Li/G) and adsorbed (Li/\mo) Li
atoms. Such suggestion is supported by calculated Li diffusion barriers along
the \lit $\rightarrow$ \lih\ diffusion path.

%\section{Computational details}

Our calculations  were performed based on the  density functional theory (DFT),
within the PBE-GGA approach~\cite{pbe}, as  implemented in the SIESTA
code~\cite{siesta}. The Kohn-Sham orbitals were expressed by a double-zeta plus 
polarization (DZP)  basis set~\cite{sankey}. The \mo/G interface was described
by using the slab method, composed by a monolayer of \mo, a single graphene
sheet, and a vacuum region of 18~\AA. In order to minimize the lattice mismatch
effects between \mo\ and graphene ($\sim$28\%), we have considered a surface (or
interface) periodicity of 7$\times$7 and 9$\times$9 for the \mo\ and graphene,
respectively. In this case, the lattice mismatch reduces to $\sim$0.6\%. All the
atomic positions are relaxed within a force convergence criteria of 10~meV/\AA.

%\section{Results and Comments}

Initially we examine the energetic stability and the equilibrium geometry of the
\mog\ interface, composed by a monolayer of \mo\ attached to the graphene sheet.
The \mog\ binding energy ($E^b$) was obtained by comparing the the total
energies of the isolated systems, \mo\ and \g\ ($E[{\rm MoS_2}]$ and $E[{\rm
G}]$, respectively), and the total energy of the \mog\ (final) system, $E[{\rm
MoS_2/G}]$, $ E^b = E[{\rm MoS_2}] + E[{\rm G}] - E[{\rm MoS_2/G}]. $ We find
that the formation of \mog\ system is an exothermic process, with $E^b$ of
21~meV/C-atom, and \mo\--G the equilibrium distance ($h$) of 3.66~\AA. Those
results are in agreement with the recent studies performed by Ma {\it et
al.}~\cite{maNanoscale2011}.
% Thus, supporting the recent confirm the weak interaction between the graphene
% sheet and the \mo~\cite{shiNanoLett2012}. Those results are in good agreement
% with recent DFT calculations, based on plane wave basis set (to expand the
% Kohn-Sham orbitals) and local density approximation (for the exchange
% correlation energy), performed by Ma {\it et al.}~\cite{maNanoscale2011}. 
Here we confirm the experimentally verified weak interaction between the
graphene sheet and the \mo~\cite{shiNanoLett2012}. Indeed, similar weak
interaction has been verified for graphene adsorbed  on SiO$_2$ or
HfO$_2$~\cite{ishigamiNanoLett2007,songNanotech2010,miwaAPL2011,scopelPRB2013}.
We are aware that DFT-GGA calculations underestimate  the binding energy of
systems ruled by vdW interactions~\cite{rydbergPRL2003}, however, we believe
that the present description of the \mog\ host structure is suitable enough for
our purposes in the present study.

Motivated by the recent experimental findings, aiming the development of LIBs
based on stacked layers of
\mo~\cite{matteAngewChem2010,duChemComm2010,hwangNanoLett2011}, the adsorption
of Li atoms on the \mo\ surface, and Li incorporation into the \mo\ bulk, have
been the subject of recent {\it ab initio} total energy
studies~\cite{liJPhysChemLett2012,chenIJElectrochemSci2013}. Here, the energetic
stability of Li adatoms on the \mo\ sheet was examined based on the calculation
of the Li binding energy ($E^b_{\rm Li}$),
$$
E^b_{\rm Li} = E[{\rm Li}] + E[{\rm MoS_2}] - E[{\rm Li/MoS_2}].
$$
$E[{\rm Li}]$ and $E[{\rm MoS_2}]$ represent the total energies of an isolated
Li atom and the pristine \mo\ monolayer, respectively, and $E[{\rm Li/MoS_2}]$
represents the total energy of the Li adsorbed \mo. We have considered Li
adatoms on the top (T) and hollow (H) sites of the \mo\ monolayer surface
[Fig.~\ref{models}(b)]. We obtained $E^b_{\rm Li}$ of 1.95 and 1.84~eV for a Li
adatom lying on the T site (\lit) and H site (\lih), respectively. The
binding energy difference ($\Delta E^b_{\rm Li}$) indicates that the \lit\
configuration is more stable than \lih\ by 0.11~eV.  At the
equilibrium geometry, the \lit\  adatom lies directly above one Mo atom, being
three-fold coordinated with the nearest neighbor S atoms. The
Li--S equilibrium bond length ($d_{\rm LiS}$) is 2.46~\AA, with a vertical
distance ($h$) of 1.57~\AA. On the H site, center of the hexagonal hole of \mo,
the \lih\ adatom also form three Li--Si bonds, with $d_{\rm LiS}$ = 2.48~\AA\
and $h$ = 1.62~\AA. The calculated Li binding energies, and the equilibrium
geometries are in good agreement with the recent DFT calculation performed by Li
{\it et al.}~\cite{liJPhysChemLett2012}. The calculated projected density of
states (PDOS) of the \lit\ adatom, Fig.~\ref{models}(c), confirms the chemical
interaction between the Li adatom and the \mo\ sheet. Through the integration of
the PDOS, we find an electronic charge transfer of 0.42 electrons from the Li
adatom to the \mo\ sheet. 

\begin{figure}[h]
%%%
% Fig 2
%%%
\includegraphics[width= 7cm]{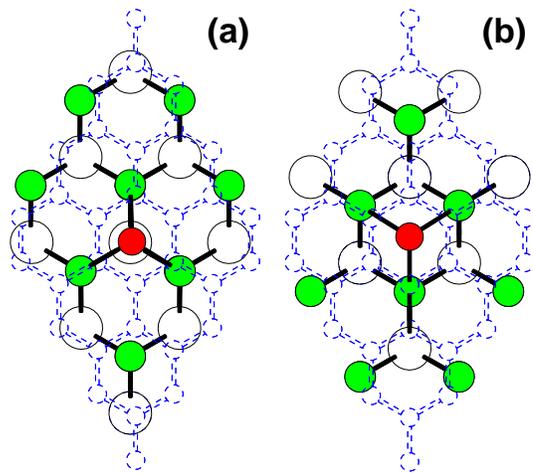}
\caption{Structural models of the \mo/Li/G system where the intercalated Li
atom lies on the T site of the \mo\ layer and holow site of the graphene (a),
and the intercalated Li atom on the H site of the \mo\ layer and hollow site of
the graphene (b).}
\label{models2}
\end{figure}

On the graphene sheet the Li adsorption is an exothermic
process~\cite{chanPRB2008}. Here we obtained $E^b_{\rm Li}$ of 1.06~eV for a Li
adatom on the hollow site, Fig.~\ref{models}(d). At the equilibrium geometry the
Li adatom lies at 1.82~\AA\ from the graphene sheet, with $d_{\rm LiC}$ of
2.33~\AA.  In Ref.~\cite{chanPRB2008}, for the same (energetically more stable)
configuration, the authors obtained $E^b_{\rm Li}$ = 1.10~eV, and  Li vertical
distance of 1.71~\AA. For both systems, Li/\mo\ and Li/G, the atomic
displacements of the host elements (bonded to the Li adatom) are very small,
that is, the \mo\ and graphene lattice structures are weakly perturbed upon the
presence of Li adatoms. 

% Our results of (i) Li adsorption energies on \mo\ and graphene sheets,
% (ii) the energetic preference (by 0.10~eV) of T site compared with the H in
% \mo, and the equilibrium geometry of Li/\mo\ abd Li/G systems are in agreement
% with the recent {\it ab initio} DFT
% results~\cite{liJPhysChemLett2012,chenIJElectrochemSci2013,chanPRB2008}.

Once we have described the Li interaction with the separated systems, 
Li/\mo\ and Li/G, let us start our investigation of Li incorporation in \mog.
Here we have considered two configurations, Li adatoms lying on the \mo\ surface
(Li/\mog), and  Li atoms embedded in the \mog\ interface (\mo/Li/G)
[Fig.~\ref{models}(a)]. In Li/\mog\ the Li adsorption energy and the equilibrium
geometry are the same as those obtained for Li/\mo. This is somewhat
expected, since the \mo--G interaction is weak. Whereas, the \mo/Li/G system
exhibits a very different picture, where the Li atoms feel the presence of the
graphene sheet. That is, the Li binding energy and equilibrium geometry depends
on the atomic structure around the Li atom at the \mo/G interface.  Here, the
energetic stability and the equilibrium geometry   were mapped by considering a
set of 32 different configurations of Li atoms embedded in the \mo/G interface.

Our total energy results reveal that, the \mo/Li/G system is energetically more
stable than Li/\mo/G (in average) by 0.41 $\pm 0.01$~eV. Thus, indicating that
the incorporation of Li atoms at the \mo/G interface is more likely than the Li
adsorption on \mo/G or \mo\ surfaces.  For the  \mo/Li/G system we
obtained averaged binding energies ($\bar{E}^b_{\rm Li}$) of 2.36
(\lit) and 2.26~$\pm$0.01~eV (\lih). Similarly to the Li/\mo/G system, there is
an energetic preference for the \lit\ configuration by 0.10~eV
($\Delta\bar{E}^b_{\rm Li}$). In fact, we did not find any
stable or metastable configuration other than \lit\ and \lih. Those results
allow us to infer that the energetic stability of the Li atoms at the \mo/G
interface is ruled by the \mo\ sheet. Indeed, the calculated Li binding
energies of the separated systems, namely Li/G (1.06~eV) and Li/\mo\ (1.95 and
1.84~eV), indicate a stronger interaction between the Li adatom and the \mo\
surface. The Li--S equilibrium bond length  is slightly stretched when compared
with the Li/\mo/G system, we find $d_{\rm LiS}$ between 2.49 and 2.61~\AA, and
vertical distances ($h$) between 1.64 and 1.77~\AA. Similarly to  Li/\mo,
there is a chemical interaction between the Li atom and the \mo/G host,
Fig.~\ref{models}(c), however, the net electronic charge transfer is negligible
in \mo/Li/G.

Although the dominant role played by the \mo\ layer, the influence
of the graphene sheet, on the energetic properties of \mo/Li/G, can be verified
by considering two particular geometries, {\it viz.}:
% due to the incommensurable charater relative position of the graphene sheet
% the \mo\ monolayer, the \mo/Li/G system exhibits a large variety of
% configurations around the intercalated Li atoms. In particular there are two
% \mo/Li/G configurations, Fig.~\ref{models2},  that worth to be examined. 
(i) the intercalated Li atom lying on the T site of \mo\ (\lit), aligned with
the hollow site of graphene [Fig.~\ref{models2}(a)], and (ii) the intercalated
Li atom lying on the H site of \mo, aligned with the hollow site of graphene
[Fig.~\ref{models2}(b)]. In (i) the \lit\ interaction with the \mo/G
interface is strengthened, and $E^b({\rm Li})$ increases to 2.40~eV, while
the binding energy of \lih\ reduces slightly to
2.24~eV, both compared with the averaged values of $\bar{E}^b_{\rm Li}$, 2.36
and 2.24~eV, respectively. For this particular configuration, the total energy
difference between
\lit\ and \lih, $\Delta E^b_{\rm Li}$,  increases to 0.16~eV.  Meanwhile, in
(ii) the binding energy of \lih\ increases to 2.33~eV, becoming the same as that
for \lit, $\Delta E^b_{\rm Li} \approx$  0. Those findings allow us to infer
that in (i), the Li atomic jumping rate from the T to H sites
(\lit~$\rightarrow$\lih) will be reduced, while in (ii) \lit~$\rightarrow$\lih\
will be increased. 

\begin{figure}[h]
%%%
% Fig 3
%%%
\includegraphics[width= 7cm]{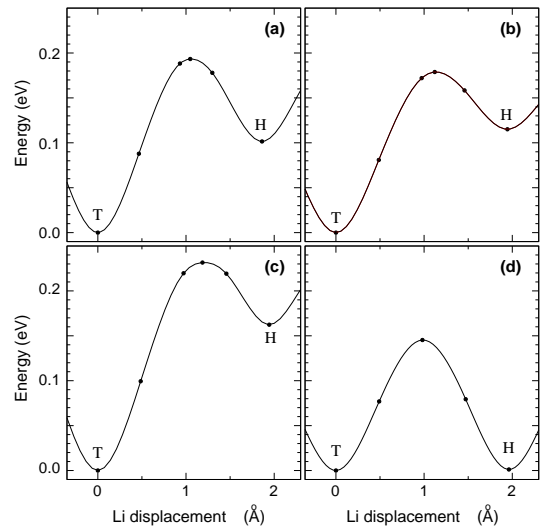}
\caption{Energy barriers for the \lit\ $\rightarrow$ \lih\ diffusion path of Li
adsorbed Li/\mo\ surface (a), and intercalated \mo/Li/G system with
$\Delta\bar{E}^b_{\rm Li}$ = 0.10~eV (b), $\Delta{E}^b_{\rm Li}$ = 0.16~eV (c),
and  $\Delta{E}^b_{\rm Li} \approx$ 0 (d).}
\label{barr2}
\end{figure}

In order to get a more complete picture of Li diffusion in the \mo/Li/G system,
we estimate the \lit~$\rightarrow$~\lih\ diffusion barrier. Here, along the 
\lit\ $\rightarrow$ \lih\ diffusion path, the Li energy barrier was obtained by
keeping fixed the Li coordinates parallel to the \mo/G interface, while both the
Li coordinate perpendicular to the interface and all the other atomic positions
were fully relaxed. By using such approach we find a diffusion barrier of
0.19~eV for Li adatom on the \mo\ monolayer surface, Fig.~\ref{barr2}(a), in
good agreement with Ref.~\cite{liJPhysChemLett2012}. For the particular
configurations discussed above, the diffusion barrier increases to 0.23~eV in
(i) [Fig.~\ref{barr2}(b)], while it reduces to 0.15~eV in (ii)
[Fig.~\ref{barr2}(d)]. Meanwhile, for the most common configurations
($\Delta\bar{E}^b_{\rm Li} = 0.10$~eV) we find an Li diffusion barrier of around
0.18~eV, similar to the one obtained for the Li/\mo\ system,
Figs.~\ref{barr2}(b) and \ref{barr2}(a), respectively. Thus, we can infer that
the diffusion barriers of Li atoms intercalated in the \mo/G interface are (in
average) similar to the diffusion barrier verified for Li adatoms on the \mo\
monolayer surface. In a recent experimental work~\cite{duChemComm2010}, the
authors verified that the mobility of Li atoms in \mo\ bulk can be enhanced by
increasing the interplanar (stacking) distance between the \mo\ layers. Indeed,
the total energy calculation performed by Li {\it et
al.}~\cite{liJPhysChemLett2012} support such statement. They find that the (\lit
$\rightarrow$ \lih) energy barrier reduces from 0.49~eV in \mo\ bulk to  0.21~eV
in \mo\ surface. The higher diffusion barrier of Li can be attributed
to the formation of Li--S chemical bonds at the interlayer regions of the \mo\
bulk. Within this scenario, \mo/Li/G represents a very interesting/promising
system for LIBs. The Li atoms are energetically more stable at the \mo/G
interface, compared with Li adatoms on the \mo\ surface, and  due to the larger
\mo--G vertical distance, and weak interaction between the graphene sheet and
the Li atoms, there is no significant increase on the Li diffusion barrier,
when compared with the Li/\mo\ system. Indeed by summing the vertical distances,
$h$, of Li/\mo\ ($\sim$1.6~\AA) and Li/G (1.7~\AA) systems, we find that it is
smaller than the equilibrium vertical distance between the graphene sheet and
the \mo\ monolayer (3.66~\AA), even by including the vdW interaction to describe
the \mo/G interface~\cite{vdW}. In this case, we find a \mo\--G equilibrium
vertical distance of 3.40~\AA.

%\section{Conclusions}

In summary, based on {\it ab initio} calculations, we examined the energetic
stability of Li atoms embedded in the \mo/G interface. Our total energy results 
indicate  that there is an energetic preference for Li atoms intercalated in the
\mo/G interface  (\mo/Li/G), when compared with Li adatoms on the \mo\ surface
(Li/\mo\ and Li/\mo/G). Thus, indicating  an increase on the Li storage capacity
at the \mo/G interfaces, in comparison with Li adatoms on the \mo\ surface. The
equilibrium geometries of the Li intercalated systems are ruled by the \mo\
layer, where the Li atoms lie on the T (\lit) and H (\lih) sites of the \mo.
Although the Li atoms feel the presence of the graphene sheet, in average we
find similar  \lit $\rightarrow$ \lih\ diffusion barriers for the \mo/Li/G and
Li/\mo\ systems. Here we can infer that, in addition to the
increase on the Li storage capacity, there is no reduction on the mobility of
the intercalated Li atoms, when compared with the one for Li adatoms on the \mo\
surface. Those findings reveal that \mo/G interfaces are interesting/promising
systems to the development of Li batteries based on \mo.

\acknowledgments

The authors acknowledge financial support from the Brazilian agencies
CNPq/INCT, and FAPEMIG,  and the computational support from CENAPAD/SP.

%\bibliography{/home/hiroki/Trab/RHMiwa}

\end{document}